\documentclass[aps,prb,superscriptaddress,amsmath,amssymb,twocolumn,10pt]{revtex4-2}
\usepackage{graphicx}
\usepackage{svg}
\usepackage{bm}
\usepackage{soul}
\usepackage{color}
\usepackage{array}
\usepackage{physics}
\usepackage[colorlinks=true, allcolors=blue]{hyperref}%
  \usepackage{times,txfonts}
  \usepackage{gensymb}
  \usepackage{siunitx}
  \usepackage{dcolumn}
  \usepackage{multirow}
  \usepackage{bm}
  \usepackage{makecell}
  \usepackage{booktabs}
  \usepackage[normalem]{ulem}
  
\begin{document}
\title{Electronic structures of crystalline and amorphous GeSe and GeSbTe compounds using machine learning empirical pseudopotentials}

\author{Sungmo Kang}
\affiliation{School of Computational Sciences, Korea Institute for Advanced Study, Seoul 02455, Korea}
\author{Rokyeon Kim}
\email{Email: rrykim@gmail.com}
\affiliation{School of Computational Sciences, Korea Institute for Advanced Study, Seoul 02455, Korea}
\author{Seungwu Han}
\email{Email: hansw@snu.ac.kr}
\affiliation{Center for AI and Natural Sciences, 
Korea Institute for Advanced Study, Seoul 02455, Korea}
\affiliation{Department of Materials Science and Engineering and Research Institute of Advanced Materials, Seoul National University, Seoul 08826, Korea}
\author{Young-Woo Son}
\email{Email: hand@kias.re.kr}
\affiliation{School of Computational Sciences, Korea Institute for Advanced Study, Seoul 02455, Korea}
\affiliation{Center for AI and Natural Sciences, 
Korea Institute for Advanced Study, Seoul 02455, Korea}
\date{\today}

\begin{abstract}
The newly developed machine learning (ML) empirical pseudopotential (EP) method overcomes the poor transferability of the traditional EP method with the help of ML techniques while preserving its formal simplicity and computational efficiency. We apply the new method to binary and ternary systems such as GeSe and Ge-Sb-Te (GST) compounds, well-known materials for non-volatile phase-change memory and related technologies. Using a training set of {\it ab initio} electronic energy bands and rotation-covariant descriptors for various GeSe and GST compounds, we generate  transferable EPs for Ge, Se, Sb, and Te. We demonstrate that the new ML model accurately reproduces the energy bands and wavefunctions of structures outside the training set, closely matching first-principles calculations. This accuracy is achieved with significantly lower computational costs due to the elimination of self-consistency iterations and the reduced size of the plane-wave basis set. Notably, the method maintains accuracy even for diverse local atomic environments, such as amorphous phases or larger systems not explicitly included in the training set.
\end{abstract}

\maketitle

\section{Introduction}
In condensed matter physics and materials science, electronic structure calculations are essential to understand the physical properties of solids and molecules. First-principles calculation methods based on density functional theory (DFT) are widely considered effective tools to obtain the quantum mechanical properties of materials~\cite{Hohenberg1964PRB,Kohn1965PR,Jones2015RMP}.  
However, DFT calculations are computationally demanding due to the iterative process required to solve the Kohn-Sham equations self-consistently~\cite{Jones2015RMP}. In addition, although there has been significant progress in linear scaling DFT calculations~\cite{Goedecker1999RMP, Soler2002JPC,Skylaris2005JCP, Ozaki2018PRB,  Nakata2020JCP, Dogan2023JCP}, computational time still increases significantly with system size and improved accuracy, posing challenges for materials simulations involving a large number of systems. 

In recent decades, machine learning (ML) techniques have been intensively developed, playing a crucial role in physics, materials science, and chemistry using data obtained from DFT calculations~\cite{Huang2023Scince,Fiedler2022PRM,Ryczko2019PRA,Duan2021JPCL,del_Rio2023npjCM,Nelson2019PRB,Pereira2017JCIM,Nagai2020npjCM,Pederson2022NRP,Schmidt2017CM,Riemelmoser2023JCTC,Brockherde2017NC,Li2024PRL}. 
Due to ongoing advancements in the refinement of techniques for more accurate and efficient predictions of physical quantities such as total energy and interatomic potentials, the combination of DFT with ML is poised to become a transformative tool in the field of material design~\cite{Schmidt2019npjCM, Schleder2019JCM, Kulik2022ES,Friederich2021NM,Batra2021NatRevMat,Merchant2023Nature}. This integration could open the door to the rapid development of novel materials by providing deeper insights into their properties and behavior, accelerating innovation in areas ranging from energy storage to semiconductors~\cite{Schmidt2019npjCM, Schleder2019JCM, Kulik2022ES,Friederich2021NM,Batra2021NatRevMat,Merchant2023Nature}. With the power of ML, DFT-based simulations could not only enhance predictive capabilities, but also streamline the discovery process~\cite{Schmidt2019npjCM, Schleder2019JCM, Kulik2022ES,Friederich2021NM,Batra2021NatRevMat,Merchant2023Nature}.

Recently, several new methodologies have emerged that integrate ML techniques with DFT to compute and predict electronic properties in condensed matter systems~\cite{Li2022NCS,Gong2023NC, Li2023NCS, Wang2024SB,  Grisafi2019ACSCS, Chandrasekaran2019npjCM,Lewis2021JCTC, Chen2021JCTC,Leonard2021JCP, Pathrudkar2024npjCM, Kim2024PRB}. 
One approach involves utilizing a neural network to represent the tight-binding (TB) Hamiltonian matrix, enabling  a direct mapping from atomic information to the matrix elements of TB Hamiltonian for a given material. Additionally, the use of equivariant graph neural networks offers more transferable and efficient predictions of electronic properties across various systems~\cite{Li2022NCS,Gong2023NC, Li2023NCS, Wang2024SB}.

Another strategy focuses on writing Kohn-Sham (KS) Hamiltonians directly by predicting their density or local potentials using various ML frameworks~\cite{ Grisafi2019ACSCS, Chandrasekaran2019npjCM,Lewis2021JCTC, Chen2021JCTC,Leonard2021JCP, Pathrudkar2024npjCM, Kim2024PRB}. Among them, a recently developed ML method~\cite{Kim2024PRB} adopts this perspective to obtain universal empirical pseudopotentials that represent the sum of all local potentials for arbitrary atomic geometries. 
The empirical pseudopotential method (EPM)
had been widely used as an efficient tool to compute electronic structures before advent of DFT~\cite{Phillips1958PR,Brust1962PRL,Brust1962aPRL, Cohen1965PR, Cohen1966PR}.
The newly developed machine learning EPM (ML-EPM)~\cite{Kim2024PRB} overcomes the poor transferability known 
for previous EPMs~\cite{Yang1974SSC, Chelikowsky1976PRB, Yeh1994PRB, Mader1994PRB, Zunger1996ASS}
and significantly reduces computational cost by bypassing the iterative self-consistent steps required for DFT without compromising its accuracy. However, its advantages have thus far been demonstrated primarily in simple crystalline systems such as silicon and silicon oxides, underscoring the need for broader testing to validate its reliability.

In this study, we extend the application of ML-EPM to a wider range of materials, focusing on binary GeSe and ternary germanium-antimony-telluride (Ge$_2$Sb$_2$Te$_5$; GST) compounds. These materials are promising candidates for phase-change memory technologies, which exploit the differences in electrical resistivity between the crystalline and amorphous phases~\cite{Ali2022NR,Liu2019JAC,Gu2022SM,Bosse2014APL,Hegedüs2008NM,Lama2020MR,Noori2021ACS,Boniardi2015ME}. However, accurately capturing the electronic properties of amorphous structures is a challenge for standard DFT calculations because of their inherent complexity and large system size. To address this, we trained an ML model based on ML-EPM and validated its performance by comparing the electronic band structures of untrained crystalline materials with results from DFT calculations. We further applied the model to amorphous structures, demonstrating its capability to produce accurate density-of-states (DOSs) predictions with significantly reduced computational costs compared to DFT calculations. Our results highlight ML-EPM as a precise and efficient framework for predicting electronic properties with excellent transferability and scalability.

\section{Formalism of ML-EPM}
\label{sec:ML-EPM}
Within a simple approximation such as the Hartree mean-field approximation for electrons in solids, the single-particle Hamiltonian of a many-atom system  
can be written as a sum of one-electron Hamiltonians, 
\begin{align}
{\mathcal H}({\bf r},\{{\bf R}_n\})=\frac{{\bf p}^2}{2m}+V({\bf r},\{{\bf R}_n\}),
\end{align}
where ${\bf r}$ and ${\bf p}$ are position and momentum of an electron and $\{{\bf R}_n\}$ a set of positions of atomic cores. 
Assuming frozen cores, i.e., $V({\bf r})\equiv V({\bf r},\{{\bf R}_n\})$, the crystal potential of $V({\bf r})$ can be further
assumed to be described by a linear superposition of atomic potentials of $V_a$ such that, 
\begin{align}
    V({\bf r})=\sum_{{\bf R},{\boldsymbol \tau}}V_a({\bf r}-{\bf R}-{\boldsymbol\tau}),
\end{align}
where ${\bf R}(\boldsymbol \tau)$ is a lattice (basis) vector.
The crystal potential can be expanded in the reciprocal lattice (its vectors are $\{\bf G\}$), 
\begin{align}
\label{form_factor}
    V({\bf r})=\sum_{\bf G}V_a({\bf G})S({\bf G})e^{i{\bf G}\cdot{\bf r}},
\end{align}
where $V_a({\bf G})$ and $S({\bf G})$ are the form factor of the atomic potential and structure factor, respectively.  
For a given crystal, ${\bf G}$, $S({\bf G})$, and $\boldsymbol \tau$ are fixed. 
So, if $V_a({\bf G})$ is known, one can compute the single-particle energy bands and other physical quantities of the given crystal under certain approximate conditions. 
The original EPM attempted to fit the atomic form factor of $V_a({\bf G})$ to experiment data~\cite{Cohen1989Springer}, thus calling the resulting local potential of $V({\bf r})$ with fitted $V_a$ empirical pseudopotential (EP)~\cite{Cohen1989Springer}. It enables efficient calculations of electronic structures or wavefunctions by adjusting the potential in the single particle Schr\"{o}dinger equation to reproduce experimental energy band splittings~\cite{Phillips1958PR,Brust1962PRL,Brust1962aPRL, Cohen1965PR, Cohen1966PR,Yang1974SSC, Chelikowsky1976PRB, Yeh1994PRB, Mader1994PRB, Zunger1996ASS}.

In DFT using the (semi)local approximation for exchange-correlation functionals, the energy band structures can be obtained by solving the KS equation self-consistently,
\begin{align}
\label{KS_equation}
\left[ -{1 \over 2}{\nabla}^2 + V_{\rm{ext}}({\bf r}) + V_{\rm{HXC}}({\bf r};n({\bf r})) \right] {\psi_i}({\bf r})={\varepsilon_i}{\psi_i}({\bf r})
\end{align}
where $V_{\rm{ext}}({\bf r})$ is the external potential, $V_{\rm{HXC}}({\bf r};n({\bf r}))$ is the sum of the Hartree and exchange-correlation potentials with electron density $n({\bf r})=\sum_i|\psi_i({\bf r})|^2$, $\varepsilon_i$ is the energy eigenvalue, and ${\psi_i}(\bold{r})$ is the electronic wavefunction~\cite{Hohenberg1964PRB,Kohn1965PR,Jones2015RMP}.
Most of the computational time is consumed by the self-consistent iteration process for $V_{\rm{HXC}}$ in Eq.~\ref{KS_equation}. Once the convergence condition is achieved, Eq.~\ref{KS_equation} can be rewritten as follows,
\begin{align}
\label{KS_equation_converge}
\left[ -{1 \over 2}{\nabla}^2 +  V_{\rm{nonlocal}} + V_{\rm DFT}({\bf r}) \right] {\psi_i}({\bf r})={\varepsilon_i}{\psi_i}({\bf r})
\end{align}
where $V_{\rm{nonlocal}}$ is the nonlocal part of the external potential $V_{\rm{ext}}(\bold{r})$, and $V_{\rm{DFT}}(\bold{r})$ is the local potential, which corresponds to a combination of the local part of the external potential, the Hartree and exchange-correlation potentials~\cite{Wang1995PRB, Fu1997PRB}.
Then, instead of adjusting the local potential to experiment, the DFT-based EPM used the converged DFT calculation results of $V_{\rm DFT}({\bf r})$ to obtain $V_a({\bf G})$ in Eq.~\ref{form_factor} or to construct $V_{\rm EP}({\bf r})$ by matching itself to $V_{\rm DFT}({\bf r})$~\cite{Wang1995PRB,Fu1997PRB}.
Although this local-density-derived EPM~\cite{Wang1995PRB} utilizing the spherical approximation indeed accelerates the calculations, the potentials of this approach still lack transferability to various materials, particularly for anisotropic crystal structures~\cite{Fu1997PRB}.

To overcome this difficulty, in the process of ML-EPM, information on the lattice and chemical environment around each atomic site is encoded in reciprocal lattice vectors of $\bf{G}$ as well as hybrid local atomic descriptors. 
The smooth overlap of atomic positions (SOAP) descriptor based on atomic neighboring density is adopted as one descriptor~\cite{Bartok2013PRB} and additionally, a rotation-covariant density coefficient descriptor is also employed to reflect directional dependence of EPs~\cite{Kim2024PRB}. These atomic descriptors and reciprocal lattice vectors are used as inputs to the ML model, where the outputs of neural network are atomic EPs.
Each atomic EP is incorporated to construct the local potential in momentum space via the following relation,
\begin{align}
V_{\mathcal C}^{\rm{ML-EPM}}(\textbf{G}_{\mathcal C})=\sum_{\alpha}S^{\alpha}_{\mathcal C}(\textbf{G}_{\rm C})v[d_{\rm C}^\alpha] (\textbf{G}_{\mathcal C})
\end{align}
where $V_{\mathcal C}^{\rm{ML-EPM}}(\textbf{G}_{\mathcal C})$ is the local potential for a given structure ${\mathcal C}$ in momentum space, $\bold{G}_{\mathcal C}$ is the reciprocal lattice vector, $S_{\mathcal C}^{\alpha} (\textbf{G}_{\mathcal C} )= {1 \over \Omega_{\mathcal C}} {\rm e}^{- {\rm i}\textbf{G}_{\mathcal C}\cdot{\boldsymbol\tau}_\alpha}$ is the structure factor of atom $\alpha$ , $\Omega_{\mathcal C}$ is the unit cell volume, and $\boldsymbol\tau_{\alpha}$ is the position vector of atom $\alpha$.
$v[d_{\mathcal C}^\alpha] (\textbf{G}_{\mathcal C})$ is the atomic EP, which depends on the descriptor $d_{\mathcal C}^\alpha$ and reciprocal lattice vector $\textbf{G}_{\mathcal C}$.
Neural networks are optimized by minimizing the loss function $L$, defined by the difference between crystal potentials obtained by ML-EPM and DFT as follows, 
\begin{align}
\label{eq:loss}
L = {1 \over N}\sum_{{\mathcal C}, \bold{G}_{\mathcal C}}\left|V^{\rm{ML-EPM}}_{\mathcal C}(\textbf{G}_{\mathcal C})-V^{\rm{DFT}}_{\mathcal C}(\textbf{G}_{\mathcal C})\right|^2
\end{align}
where $N$ is the total size of crystal potentials in the training dataset. The trained ML model produces $V_{\mathcal C}^{\rm{ML-EPM}}(\textbf{G}_{\mathcal C})$, which corresponds to the Fourier transformation of $V_{\rm{EP}}(\bold{r})$ so that one can obtain the complete KS Hamiltonian for a given crystal structure ${\mathcal C}$. The detailed architecture of the current ML model was introduced in our previous study~\cite{Kim2024PRB}. Since we obtain transferable local potentials or KS Hamiltonian directly from ML, we expect that results can be applied to various post-processing computational tools within existing first-principles calculations package as was demonstrated for calculations of frequency dependent dielectric functions of silicon oxides with an oxygen vacancy~\cite{Kim2024PRB}.

\begin{figure*}
	\includegraphics[width=1.0\linewidth]{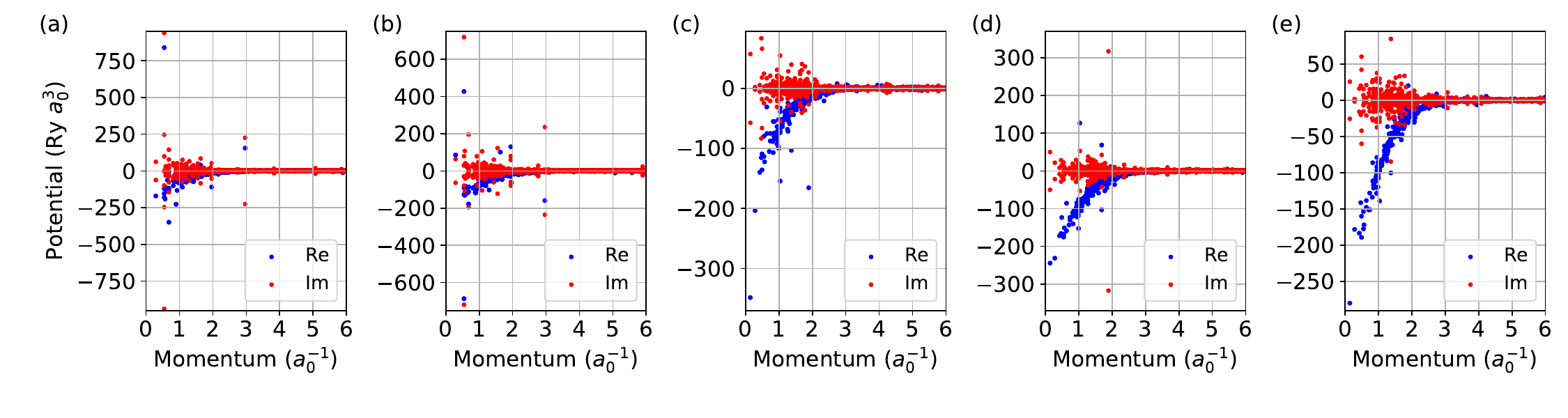}
    \caption{Momentum-dependent atomic empirical pseudopotential calculated by DFT for (a) Ge atoms, (b) Se atoms in GeSe (mp-1190257 supercell structure), and (c) Ge atoms, (d) Sb atoms, and (e) Te atoms in GeSbTe (mp-1224415 supercell structure) compounds. The mp-1190257 and mp-1224415 supercells are chosen for GeSe and GST, respectively. Blue and red circles in the plot indicate the real and imaginary parts of the atomic empirical pseudopotential, respectively. $a{_0}$ is the Bohr radius.}
	\label{fig:vloc_G}
\end{figure*}

\section{Computational Details}
To generate the training dataset, Quantum ESPRESSO code is employed for first-principles electronic structure calculations\cite{Giannozzi2009JPCM,QuantumEspresso}. Norm-conserving pseudopotentials are adopted with 4$s$, 4$p$ orbitals as the valence state for Ge and Se atoms, and 5$s$, 5$p$ orbitals as the valence state for Sb and Te atoms, respectively\cite{Hamann1979PRL}. We choose the generalized gradient approximation (GGA) for exchange-correlation functional with the parameterization of Perdew, Burke, and Enzerhof (PBE)\cite{Perdew1996PRL}. We also used a Monkhorst-Pack grid with automatically generated k-grids depending on the unit cell size of each crystal structures in the training dataset~\cite{Monkhorst1976PRB}. The kinetic energy cutoff and self-consistent convergence criterion are set to 40 Ry and 10$^{-6}$ Ry, respectively.

We follow the previous ML-EPM framework to construct a neural network for GeSe and GST compounds~\cite{Kim2024PRB}. To prepare the training dataset, crystal structures are generated by applying distortion to unitcells and random atomic displacements for each structure obtained from the Materials Project database~\cite{MaterialsProject}. 
By performing DFT calculations, we extract local potentials $V^{\rm{DFT}}(\textbf{G})$ for each generated structure, which are utilized for training the ML models. In the neural network, we use three fully connected hidden layers with 1024 neurons for each layer and ReLU activation functions~\cite{Behler2007PRL}. Our models are trained by minimizing the loss function defined in Eq.~\ref{eq:loss} using the Adam optimizer~\cite{Kingma2017arXiv}.

We focus on the target materials GeSe and GST compounds to verify the performance of ML-EPM. Predictions of electronic properties are made for unlearned crystalline structures of GeSe and GST compounds, as well as their amorphous structures. For this purpose, we prepare a total of 4 crystalline structures for GeSe and 10 crystalline structures for GST compounds with various compositions and system size of which initial structures are selected from the Materials Project database~\cite{MaterialsProject}. We then prepare supercells from their initial structures with a maximum of 18 atoms for GeSe and 24 atoms for GST in each unit cell, respectively. Additionally, we create structures for GeSe and GST by randomly distributing atoms within a cubic cell. These structures are included in the dataset to enhance prediction accuracy by representing a broader range of atomic environments beyond just crystalline structures. Including all these variations, we generate a total of 25,056 crystalline structures and 5,184 random structures for GeSe, and a total of 30,600 crystalline structures and 8,640 random structures for GST, applying a 2.5 $\sim$ 7.5\% variations in lattice constants and random atomic displacements of up to 0.5 ${\AA}$ per each atom, respectively. By performing DFT calculations for each generated structure, local potentials $V^{\rm{DFT}}(\textbf{G})$ are calculated for the dataset. The 80\% of the dataset is used for training, while the remaining 20\% is set aside as the validation set.

\section{\label{result}Results}

\begin{figure*}
	\includegraphics[width=0.8\linewidth]{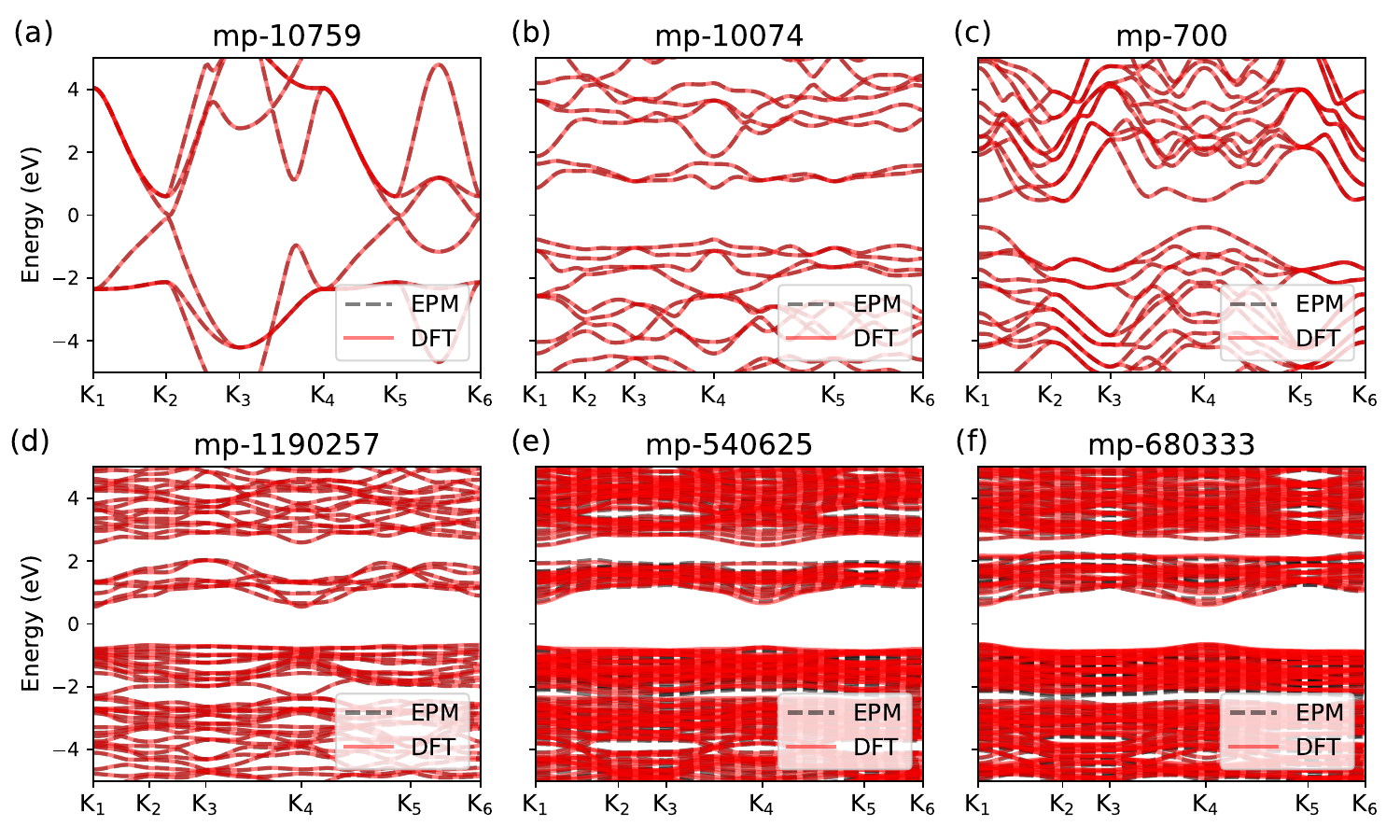}
    \caption{(a)-(f) Electronic band structures for various crystalline GeSe compounds using ML-EPM (gray dashed line) and DFT (red line). In each band structure plot, the title indicates the material ID, and the Fermi level is set to zero. ${\rm K}_i~(i = 1,\cdots, 6)$ are reduced coordinates in reciprocal space, representing $(0, 0, 0)$, $(0.5, 0 ,0)$, $(0.5, 0.5, 0)$, $(0, 0,
0)$, $(0.5, 0.5, 0.5)$, $(0.5, 0, 0)$ in their corresponding crystal structures, respectively. }
	\label{fig:band_GeSe}
\end{figure*}

Before training the ML model, for a crystal structure of $\mathcal C$, we first examine the momentum dependence of the atomic form factors of $v_{\mathcal C}^\alpha (\textbf{G}_{\mathcal C})$ to determine appropriate descriptors depending on $\bf G$. 
The atomic form factor for $\alpha$-th atom can be obtained by converting $V_{\mathcal C}(\textbf{G}_{\mathcal C})=\sum_{\alpha}S^{\alpha}_{\mathcal C}(\textbf{G}_{\mathcal C})v_{\mathcal C}^\alpha (\textbf{G}_{\mathcal C})$
where the local potential $V_{\mathcal C}(\textbf{G}_{\mathcal C})$ and structure factor $S^{\alpha}_{\mathcal C}(\textbf{G}_{\mathcal C})$ are easily obtained from standard DFT calculations.
In Fig.~\ref{fig:vloc_G}, the dependence of atomic empirical pseudopotentials $v_{\rm C}^\alpha (\textbf{G}_{\rm C})$ on the direction of reciprocal lattice vector $\bold{G}$ almost disappears when the magnitude of $\bold{G}$ increases. Therefore, we assume a spherical approximation and use the rotationally invariant SOAP descriptor when $|\textbf{G}|$ $>$ 3.0 $a{_0}^{-1}$. For $|\textbf{G}|$ $<$ 3.0 $a{_0}^{-1}$, we employ a rotationally covariant atomic density coefficient descriptor to account for directional dependencies, thus ensuring directional dependence of the bonds in learned potentials~\cite{Kim2024PRB}.

\begin{table}[b]
\caption{Root mean square errors of the SOAP descriptor ($\Delta p$) and energy eigenvalues ($\Delta E$) for GeSe crystalline structures used in Fig.~\ref{fig:band_GeSe}.}
\label{table:GeSe}
\begin{ruledtabular}
\begin{tabular}{cccc}
 \ Material ID & Stoichiometry & $\Delta p$ & $\Delta E$  (meV) \\
 \hline
   mp-10759 & GeSe & 2.850 & 6.582 \\
   mp-10074 & Ge$_{2}$Se$_{4}$  & 3.123 & 3.763 \\
   mp-700 & Ge$_{4}$Se$_{4}$  & 3.622 & 5.881 \\
   mp-1190257 & Ge$_{6}$Se$_{12}$ & 3.338 & 7.634 \\
   mp-540625 & Ge$_{16}$Se$_{32}$  & 6.951 & 50.304 \\
   mp-680333 & Ge$_{16}$Se$_{36}$  & 8.760 & 66.497 \\
\end{tabular}
\end{ruledtabular}
\end{table}

Based on these observations, we train ML models with momentum-separated datasets for GeSe and GST compounds individually. We tested our trained model by calculating the electronic band structure for different unlearned crystalline structures with various compositions and system sizes from the Materials Project database~\cite{MaterialsProject}. Fig.~\ref{fig:band_GeSe} shows the results of the electronic band structure calculation of GeSe obtained by ML-EPM and DFT calculation for comparison.

\begin{figure} [b]
	\includegraphics[width=1.0\linewidth]{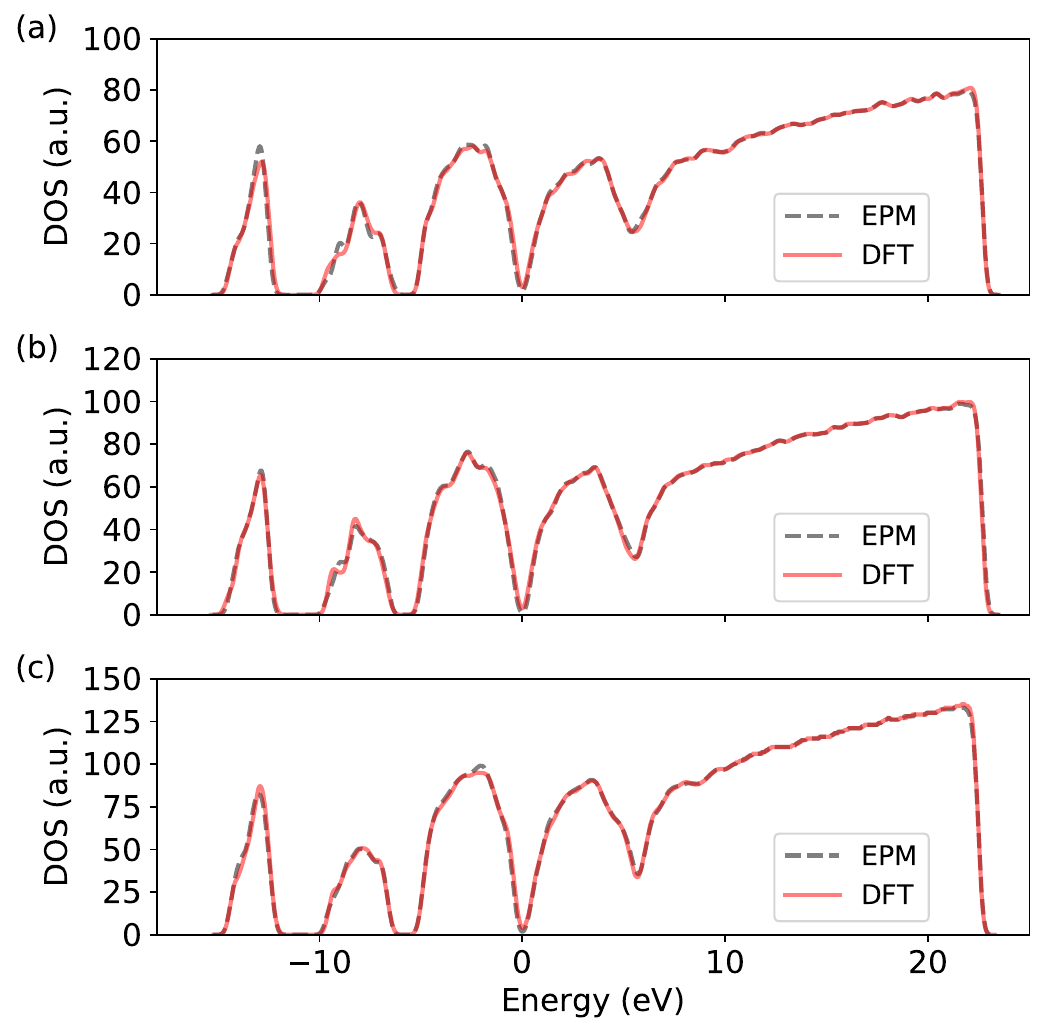}
    \caption{Total density of states as a function of energy for amorphous GeSe calculated by ML-EPM (gray dashed line) and DFT (red line) for (a) 72 atoms, (b) 90 atoms, and (c) 120 atoms in the unit cell, respectively. The Fermi level is set to zero for each plot.}
	\label{fig:DOS_GeSe}
\end{figure}

\begin{figure*}
	\includegraphics[width=0.8\linewidth]{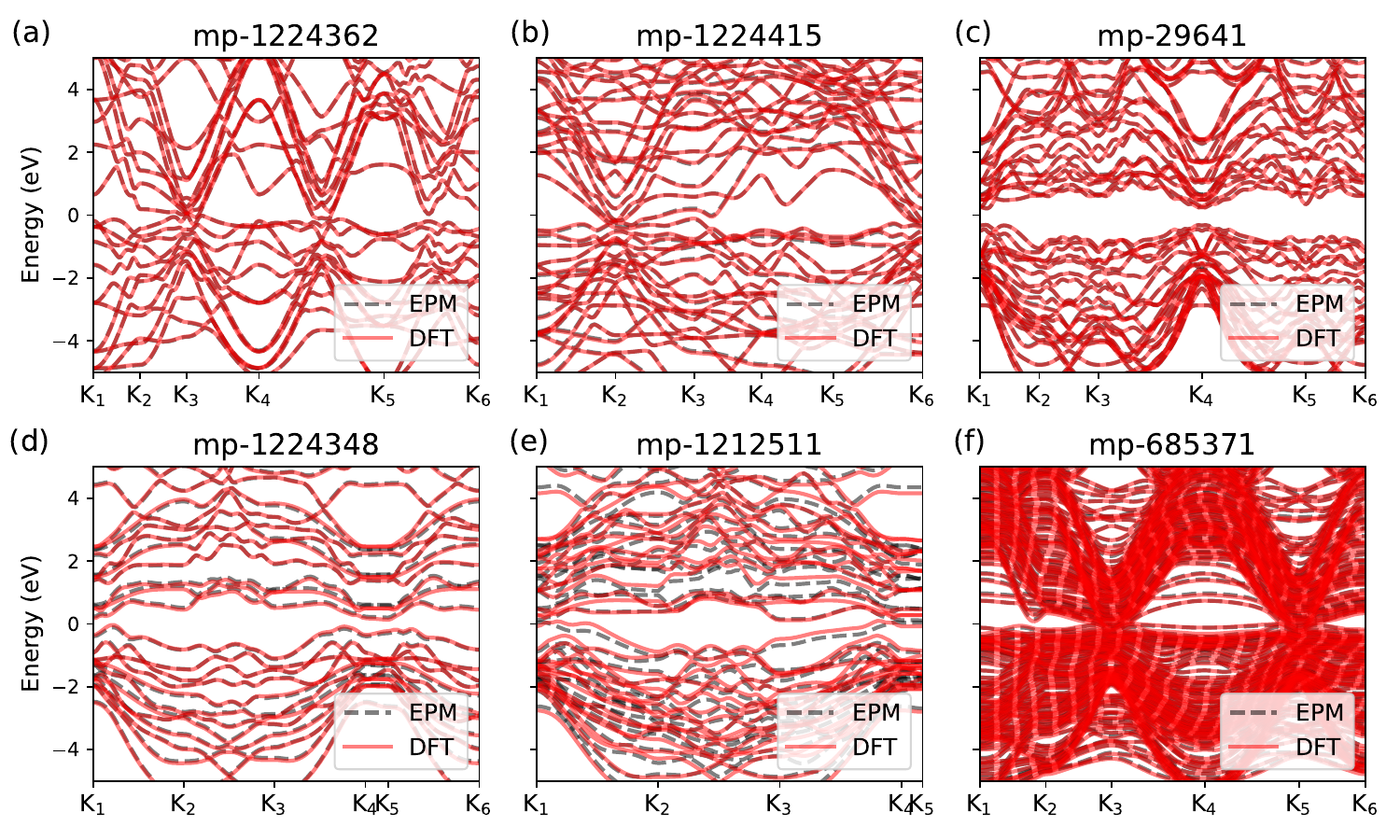}
    \caption{(a)-(f) Electronic band structures for various crystalline GeSbTe compounds using ML-EPM (gray dashed line) and DFT (red line). In each band structure plot, the title indicates material ID, and the Fermi level is set to zero. ${\rm K}_i~(i = 1,\cdots, 6)$ are reduced coordinates in reciprocal space, representing $(0, 0, 0)$, $(0.5, 0 ,0)$, $(0.5, 0.5, 0)$, $(0, 0,
0)$, $(0.5, 0.5, 0.5)$, $(0.5, 0, 0)$ in their corresponding crystal structures, respectively. }
	\label{fig:band_GST}
\end{figure*}

In all structures for test set, the ML-EPM band structures are shown to align well with DFT calculations. Interestingly, our ML models demonstrate remarkable prediction accuracy for symmetric structures, even though they were trained solely on randomly distorted structures without explicit crystal symmetry. To analyze our electronic band structure prediction results more quantitatively, we evaluate the structural differences of test structures and calculate the root mean square error of energy eigenvalues, as listed in Table ~\ref{table:GeSe}.
To quantify the structural difference from the dataset structures, we define the root mean square error of the SOAP descriptor $p$ for each atomic species $Z$ using the following equation,
\begin{align}
\Delta p = \sqrt{
{1 \over N}\sum_{Z}{\rm min}_{Z^\prime}
\left[
\sum_{m,m^\prime,l}
     \left(
       p^{Z^\prime, {\rm dataset}}_{m,m^\prime,l}-p^{Z}_{m,m^\prime,l}
    \right)^2
\right] 
}
\end{align}
where $m$, $m^\prime$, and $l$ are component indices of the SOAP descriptor $p$, and $N$ is the number of atoms in the unit cell. The root mean square error of energy eigenvalues is expressed as
\begin{align}
\Delta E = \sqrt{
{1 \over {N_e}} \sum_{n,k} 
\left(
    E^{\rm DFT}_{n,k} - E^{\rm ML-EPM}_{n,k}
\right)^2
}
\end{align}
where $n$ is the band index, $k$ the k-point along a high-symmetry line, and $N_e$ the total number of energy eigenvalues.
For $\Delta p$, a larger value signifies a stronger deviation from the learned structures. For $\Delta E$, a smaller value indicates that the predicted band eigenvalues from machine learning are closer to the DFT ones.

\begin{table}[b]
\caption{Root mean square errors of the SOAP descriptor ($\Delta p$) and energy eigenvalues ($\Delta E$) for GST crystalline structures used in Fig.~\ref{fig:band_GST}.}
\label{table:GST}
\begin{ruledtabular}
\begin{tabular}{cccc}
 \ Material ID & Stoichiometry & $\Delta p$ & $\Delta E$ (meV) \\
 \hline
   mp-1224362 & GeSb$_{2}$Te$_{4}$ & 2.348 & 8.735 \\
   mp-1224415 & Ge$_{3}$Sb$_{2}$Te$_{6}$  & 2.489 & 21.597 \\
   mp-29641 & GeSb$_{4}$Te$_{7}$  & 2.713 & 14.190 \\
   mp-1224348 & GeSb$_{2}$Te$_{4}$ & 5.591 & 52.877 \\
   mp-1212511 & Ge$_{3}$Sb$_{2}$Te$_{6}$  & 6.945 & 117.531 \\
   mp-685371 & Ge$_{9}$Sb$_{10}$Te$_{24}$  & 10.326 & 29.302 \\
\end{tabular}
\end{ruledtabular}
\end{table}

In Table~\ref{table:GeSe}, the data sets were generated by distorting the structures of mp-10759, mp-10074, mp-700, and mp-1190257, which produce low $\Delta p$ values compared to those of the structures of mp-540625 and mp-680333. Therefore, the trend shown here implies that our models learn similar atomic environments if compared with mp-10759, mp-10074, mp-700, and mp-1190257 structures, resulting in highly accurate electronic band structure predictions. However, the models also demonstrate strong predictive performance for unlearned structures like mp-540625 and mp-680333. Additionally, mp-540625 and mp-680333 structures contain 48 and 52 atoms, respectively, which is significantly larger than the maximum of 18 atoms, a criteria for generating our current dataset. We also evaluate our trained model by calculating the electronic properties of amorphous structures, which were generated using the general melt-quench method~\cite{Cho2010JPCM,Lee2020CMS,Cho2011APL}. Fig.~\ref{fig:DOS_GeSe} shows the density of states for three different amorphous structures of GeSe. Although our models were trained on distorted crystals, the predicted density of states from ML-EPM for those amorphous structures aligns with DFT results regardless of their system sizes. We also note that the number of atoms in amorphous GeSe is quite larger than our training criteria.

\begin{figure}[b]
	\includegraphics[width=1.0\linewidth]{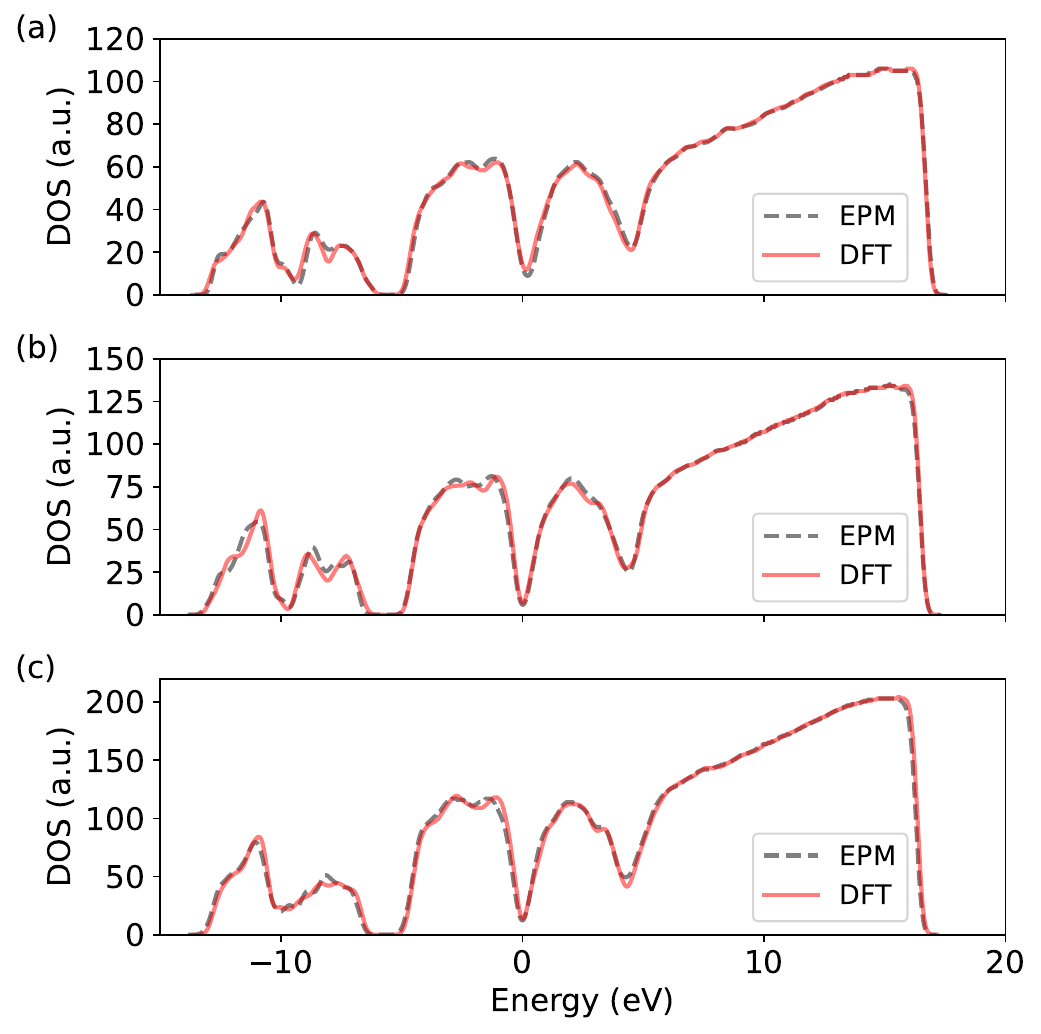}
    \caption{Total density of states as a function of energy for amorphous GeSbTe calculated by ML-EPM (gray dashed line) and DFT (red line) for (a) 72 atoms, (b) 90 atoms, and (c) 135 atoms in unit cell, respectively. The Fermi level is set to zero for each plot.}
	\label{fig:DOS_GST}
\end{figure}

Next, we applied our ML model to calculate electronic structures of GST, expanding our model's applicability to ternary compounds beyond binary systems~\cite{Kim2024PRB}. The process is similar to that for GeSe case and for silicon oxides in our previous study~\cite{Kim2024PRB}. Fig.~\ref{fig:band_GST} shows electronic band structures for GST obtained through ML-EPM and DFT calculations. For all crystalline GST test structures, the ML-EPM band structures show good agreement with corresponding DFT calculations, despite the more complex chemical environments of the ternary system.

In TABLE.~\ref{table:GST}, the root mean square errors of SOAP descriptor and energy eigenvalues for test structures are listed. For training, datasets were generated by distorting mp-1224362, mp-1224415, and mp-29641 structures of GST compounds. The models show good predictive accuracy for these similar structures, as well as for unlearned structures like mp-1224348, mp-1212511, and mp-685371. Among the unlearned structures, mp-685371 contains 43 atoms in its unit cell, which is much larger than the maximum of 24 atoms used for generating the dataset.

\begin{table}[t]
\caption{Computational time (in second) estimation for various crystalline structures of GeSe and GST with different number of atoms in unit cell ($N_a$) using DFT and ML-EPM denoted as $T_{\rm DFT}$ and $T_{\rm ML-EPM}$, respectively. All calculations are performed on a single node of a local cluster.\footnote{Intel(R) Xeon E5-2680 v3(2.5, GHz) (24 cores)}}
\label{table:computational time}
\begin{ruledtabular}
\begin{tabular}{cccc}
 \ Material ID & $N_a$ & $T_{\rm DFT}$ & $T_{\rm ML-EPM}$ \\
 \hline
   mp-10759 & 2 & 4.36 & 1.69 \\
   mp-10074 & 6 & 17.06 & 7.65 \\
   mp-700 & 8 & 16.00 & 5.75 \\
   mp-1190257 & 18 & 109.81 & 30.88 \\
   mp-540625 & 48 & 418.20 & 138.50 \\
   mp-680333 & 52 & 249.54 & 87.55 \\
   mp-1224362 & 7 & 48.76 & 21.59 \\
   mp-1224415 & 11 & 44.40 & 19.66 \\
   mp-29641 & 12 & 142.97 & 32.81 \\
   mp-1224348 & 7 & 550.58 & 175.62 \\
   mp-1212511 & 11 & 902.61 & 142.36 \\
   mp-685371 & 43 & 999.71 & 167.42 \\
\end{tabular}
\end{ruledtabular}
\end{table}

We also assess our model by calculating the electronic properties of amorphous structures. Fig.~\ref{fig:DOS_GST} shows the density of states for GST amorphous structures with various system sizes. The results show good agreements with the corresponding DFT calculations for both crystalline and amorphous structures. 
For all considered amorphous structures, spectral gaps below Fermi energy as well as suppressed density of states at specific energy levels are well captured irrespective of system sizes.  

Having demonstrated the accuracy of ML-EPM, we now present the efficiency of the model. Since our EPs have trained for converged local DFT potentials, the KS equation with $V_{\rm EP}$ replacing $V_{\rm DFT}$ in Eq.~\ref{KS_equation_converge} does not have self-consistent iteration, thus providing a chance to compute energy eigenvalues with reduced computational resources. We compare the computational time for calculating the electronic band structures of crystalline GeSe and GST using DFT and ML-EPM, as listed in Table~\ref{table:computational time}. 
Notably, as the DFT computatonal time increases, the computational time required for ML-EPM decreases dramatically. For example, for mp-10074, mp-1224415, mp-1224362, and mp-10759, our ML-EPM is about $220\sim 260~\%$ more efficient than DFT, while for mp-1212511 and mp-685371, it is about $600\%$. We also note that as the number of atoms increases, the efficiency is also enhanced. For systems with more than 40 atoms, the typical efficiency is over $300~\%$.

\begin{figure}[t]
	\includegraphics[width=1.0\linewidth]{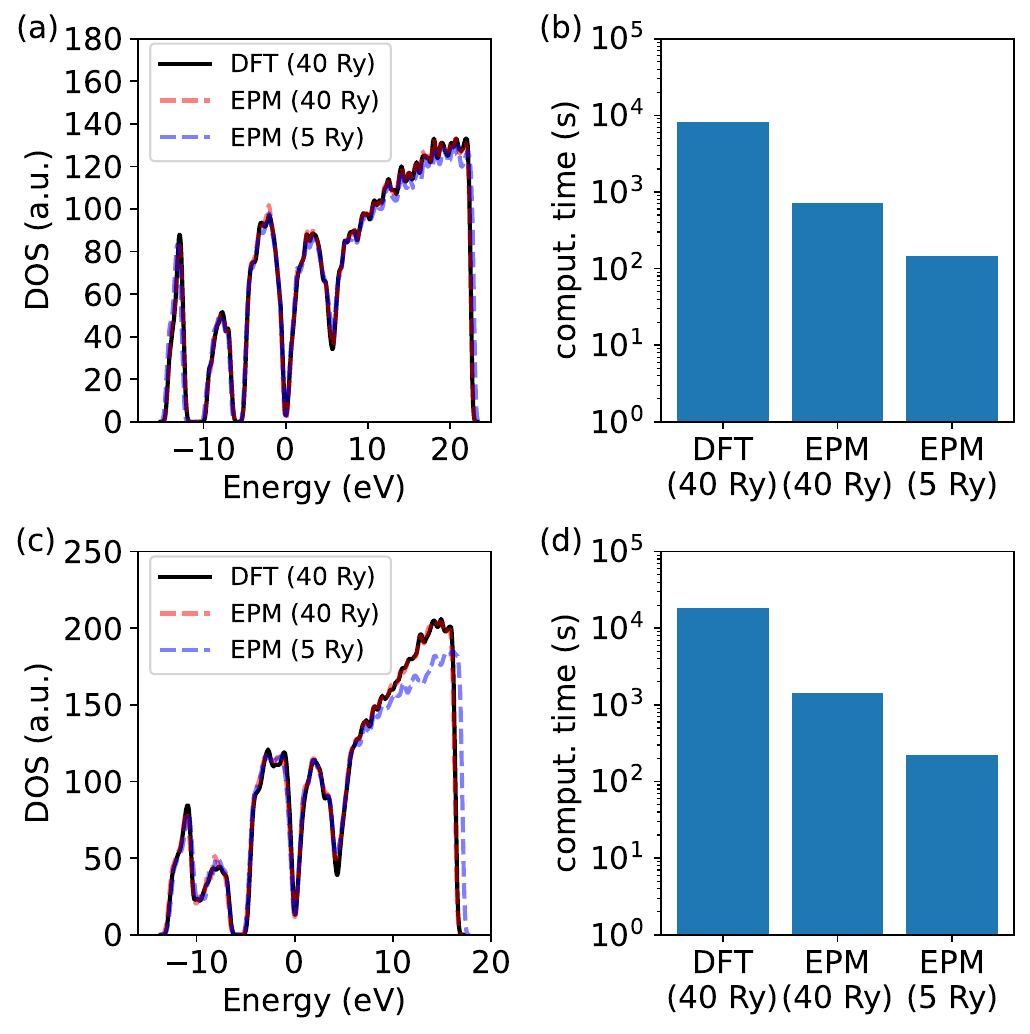}
    \caption{(a) Total density of states (TDOSs) of amorphous GeSe with 120 atoms as a function of energy obtained by DFT and ML-EPM with a cutoff energy of 40 Ry and 5 Ry, respectively. (b) Comparison of computational times for obtaining (1) from DFT and ML-EPM with different cut-off energy of 40 Ry and 5 Ry. (c) and (d)   Similar comparions of (c) TDOSs and (d) computational time for amorphous GST with 135 atoms. The DFT calculation results are denoted as black solid lines, while ML-EPM results with energy cutoff of 40 Ry and 5 Ry are indicated as red and blue dashed lines, respectively.}
	\label{fig:DOS_time}
\end{figure}

Moreover, as was demonstrated for silicon oxide before~\cite{Kim2024PRB}, the energy bands of even larger systems can be obtained quite efficiently with our ML-EPM. Since our model has already converged potentials, the energy cutoff required for expansion for plane-wave basis set is only limited by the kinetic energy in the KS Hamiltonian in Eq.~\ref{KS_equation_converge}~\cite{Kim2024PRB}. This allows for a considerable reduction in the size of the Kohn-Sham Hamiltonian matrix and the number of plane wave basis functions, further saving computational time. To confirm this, we computed energy spectrum and total density of states (TDOS)  of larger amorphous GeSe and GST compounds which have 120 and 135 atoms in their unit cells, respectively, with significantly reduced energy cutoff. As shown in Fig.~\ref{fig:DOS_time}, almost all spectral features of both compounds, including spectral gaps, peaks, and dips in the TDOS, remain consistent when the energy cutoff is reduced from 40 Ry to 5 Ry, except for states at very high energy. This reduction significantly accelerates computational speed, making it nearly 100 times faster than DFT calculations.

\section{Conclusion}
In this work, we introduced a couple of applications of a novel ML computational method that leverages neural networks to achieve accuracy in computing quantum mechanical energy bands comparable to first-principles calculation methods based on DFT. Our approach integrates the strengths of traditional empirical pseudopotential methods and ML by incorporating a hybrid descriptor and the Materials Project database that reliably captures anisotropic local structures, even in the amorphous systems. We demonstrated the effectiveness and versatility of this method by computing energy bands for various  GeSe and GST compounds, including their amorphous structures with a large number of atoms. 

Our transferable empirical pseudopotentials can replace all local Hartree, atomic and exchange-correlation potentials in KS Hamiltonians without requiring self-consistent iterations, making them compatible with existing first-principles calculation program packages. We also note that the current method is capable of learning an advanced {\it ab initio} technique ~\cite{Kim2024PRB} such as extended Hubbard functionals~\cite{Lee2020PRR, Yang2021PRB}. Furthermore, as shown in amorphous GeSe and GST structures, our methodology is well-suited for nonideal structures. This suggests its potential for generating transferable empirical pseudopotentials for molecules, clusters and other disordered or defective systems. Ultimately, we anticipate that our universal ML-EPM will play a crucial role in advancing computation-driven materials databases.  

\begin{acknowledgements}
We thank Hongkee Yoon and Minseok Moon for fruitful discussions. 
S.S. and Y.-W.S. were supported by KIAS individual Grants (No. CG031509 and No. CG092401). 
S. H. was supported by the National Research Foundation
of Korea (NRF) grant funded by the Korea government (MSIT)
(RS-2023-00247245). 
Computations were supported by the CAC of KIAS.
\end{acknowledgements}

\bibliography{reference}
 
\end{document}